%
\newskip\ttglue
\font\fiverm=cmr5
\font\fivei=cmmi5
\font\fivesy=cmsy5
\font\fivebf=cmbx5
\font\sixrm=cmr6
\font\sixi=cmmi6
\font\sixsy=cmsy6
\font\sixbf=cmbx6
\font\sevenrm=cmr7
\font\eightrm=cmr8
\font\eighti=cmmi8
\font\eightsy=cmsy8
\font\eightit=cmti8
\font\eightsl=cmsl8
\font\eighttt=cmtt8
\font\eightbf=cmbx8
\font\ninerm=cmr9
\font\ninei=cmmi9
\font\ninesy=cmsy9
\font\nineit=cmti9
\font\ninesl=cmsl9
\font\ninett=cmtt9
\font\ninebf=cmbx9
\font\twelverm=cmr12
\font\twelvei=cmmi12
\font\twelvesy=cmsy12
\font\twelveit=cmti12
\font\twelvesl=cmsl12
\font\twelvett=cmtt12
\font\twelvebf=cmbx12


\def\eightpoint{\def\rm{\fam0\eightrm}  
  \textfont0=\eightrm \scriptfont0=\sixrm \scriptscriptfont0=\fiverm
  \textfont1=\eighti  \scriptfont1=\sixi  \scriptscriptfont1=\fivei
  \textfont2=\eightsy  \scriptfont2=\sixsy  \scriptscriptfont2=\fivesy
  \textfont3=\tenex  \scriptfont3=\tenex  \scriptscriptfont3=\tenex
  \textfont\itfam=\eightit  \def\it{\fam\itfam\eightit}
  \textfont\slfam=\eightsl  \def\sl{\fam\slfam\eightsl}
  \textfont\ttfam=\eighttt  \def\tt{\fam\ttfam\eighttt}
  \textfont\bffam=\eightbf  \scriptfont\bffam=\sixbf
    \scriptscriptfont\bffam=\fivebf  \def\bf{\fam\bffam\eightbf}
  \tt  \ttglue=.5em plus.25em minus.15em
  \normalbaselineskip=9pt
  \setbox\strutbox=\hbox{\vrule height7pt depth2pt width0pt}
  \let\sc=\sixrm  \let\big=\eightbig \normalbaselines\rm}

\def\eightbig#1{{\hbox{$\textfont0=\ninerm\textfont2=\ninesy
        \left#1\vbox to6.5pt{}\right.$}}}


\def\ninepoint{\def\rm{\fam0\ninerm}  
  \textfont0=\ninerm \scriptfont0=\sixrm \scriptscriptfont0=\fiverm
  \textfont1=\ninei  \scriptfont1=\sixi  \scriptscriptfont1=\fivei
  \textfont2=\ninesy  \scriptfont2=\sixsy  \scriptscriptfont2=\fivesy
  \textfont3=\tenex  \scriptfont3=\tenex  \scriptscriptfont3=\tenex
  \textfont\itfam=\nineit  \def\it{\fam\itfam\nineit}
  \textfont\slfam=\ninesl  \def\sl{\fam\slfam\ninesl}
  \textfont\ttfam=\ninett  \def\tt{\fam\ttfam\ninett}
  \textfont\bffam=\ninebf  \scriptfont\bffam=\sixbf
    \scriptscriptfont\bffam=\fivebf  \def\bf{\fam\bffam\ninebf}
  \tt  \ttglue=.5em plus.25em minus.15em
  \normalbaselineskip=11pt
  \setbox\strutbox=\hbox{\vrule height8pt depth3pt width0pt}
  \let\sc=\sevenrm  \let\big=\ninebig \normalbaselines\rm}

\def\ninebig#1{{\hbox{$\textfont0=\tenrm\textfont2=\tensy
        \left#1\vbox to7.25pt{}\right.$}}}


\def\twelvepoint{\def\rm{\fam0\twelverm}  
  \textfont0=\twelverm \scriptfont0=\eightrm \scriptscriptfont0=\sixrm
  \textfont1=\twelvei  \scriptfont1=\eighti  \scriptscriptfont1=\sixi
  \textfont2=\twelvesy  \scriptfont2=\eightsy  \scriptscriptfont2=\sixsy
  \textfont3=\tenex  \scriptfont3=\tenex  \scriptscriptfont3=\tenex
  \textfont\itfam=\twelveit  \def\it{\fam\itfam\twelveit}
  \textfont\slfam=\twelvesl  \def\sl{\fam\slfam\twelvesl}
  \textfont\ttfam=\twelvett  \def\tt{\fam\ttfam\twelvett}
  \textfont\bffam=\twelvebf  \scriptfont\bffam=\eightbf
    \scriptscriptfont\bffam=\sixbf  \def\bf{\fam\bffam\twelvebf}
  \tt  \ttglue=.5em plus.25em minus.15em
  \normalbaselineskip=11pt
  \setbox\strutbox=\hbox{\vrule height8pt depth3pt width0pt}
  \let\sc=\sevenrm  \let\big=\twelvebig \normalbaselines\rm}

\def\twelvebig#1{{\hbox{$\textfont0=\tenrm\textfont2=\tensy
        \left#1\vbox to7.25pt{}\right.$}}}

\magnification=\magstep1
\def\firstpage{1}
\pageno=\firstpage

\font\fiverm=cmr5
\font\sevenrm=cmr7

\font\eightrm=cmr8
\font\eightbf=cmbx8
\font\ninerm=cmr9
\font\ninebf=cmbx9
\font\tenbf=cmbx10
\font\magtenbf=cmbx10 scaled\magstep1

\catcode`\@=11
%

\def\undefine#1{\let#1\undefined}
\def\newsymbol#1#2#3#4#5{\let\next@\relax
 \ifnum#2=\@ne\let\next@\msafam@\else
 \ifnum#2=\tw@\let\next@\msbfam@\fi\fi
 \mathchardef#1="#3\next@#4#5}
\def\mathhexbox@#1#2#3{\relax
 \ifmmode\mathpalette{}{\m@th\mathchar"#1#2#3}%
 \else\leavevmode\hbox{$\m@th\mathchar"#1#2#3$}\fi}
\def\hexnumber@#1{\ifcase#1 0\or 1\or 2\or 3\or 4\or 5\or 6\or 7\or 8\or
 9\or A\or B\or C\or D\or E\or F\fi}

\font\tenmsa=msam10
\font\sevenmsa=msam7
\font\fivemsa=msam5
\newfam\msafam
\textfont\msafam=\tenmsa
\scriptfont\msafam=\sevenmsa
\scriptscriptfont\msafam=\fivemsa
\edef\msafam@{\hexnumber@\msafam}
\mathchardef\dabar@"0\msafam@39
\def\dashrightarrow{\mathrel{\dabar@\dabar@\mathchar"0\msafam@4B}}
\def\dashleftarrow{\mathrel{\mathchar"0\msafam@4C\dabar@\dabar@}}

\def\ulcorner{\delimiter"4\msafam@70\msafam@70 }
\def\urcorner{\delimiter"5\msafam@71\msafam@71 }
\def\llcorner{\delimiter"4\msafam@78\msafam@78 }
\def\lrcorner{\delimiter"5\msafam@79\msafam@79 }
\def\yen{{\mathhexbox@\msafam@55}}
\def\checkmark{{\mathhexbox@\msafam@58}}
\def\circledR{{\mathhexbox@\msafam@72}}
\def\maltese{{\mathhexbox@\msafam@7A}}

\font\tenmsb=msbm10
\font\sevenmsb=msbm7
\font\fivemsb=msbm5
\newfam\msbfam
\textfont\msbfam=\tenmsb
\scriptfont\msbfam=\sevenmsb
\scriptscriptfont\msbfam=\fivemsb
\edef\msbfam@{\hexnumber@\msbfam}
\def\Bbb#1{{\fam\msbfam\relax#1}}
\def\widehat#1{\setbox\z@\hbox{$\m@th#1$}%
 \ifdim\wd\z@>\tw@ em\mathaccent"0\msbfam@5B{#1}%
 \else\mathaccent"0362{#1}\fi}
\def\widetilde#1{\setbox\z@\hbox{$\m@th#1$}%
 \ifdim\wd\z@>\tw@ em\mathaccent"0\msbfam@5D{#1}%
 \else\mathaccent"0365{#1}\fi}
\font\teneufm=eufm10
\font\seveneufm=eufm7
\font\fiveeufm=eufm5
\newfam\eufmfam
\textfont\eufmfam=\teneufm
\scriptfont\eufmfam=\seveneufm
\scriptscriptfont\eufmfam=\fiveeufm

\catcode`\@=11
\newsymbol\boxdot 1200
\newsymbol\boxplus 1201
\newsymbol\boxtimes 1202
\newsymbol\square 1003
\newsymbol\blacksquare 1004
\newsymbol\centerdot 1205
\newsymbol\lozenge 1006
\newsymbol\blacklozenge 1007
\newsymbol\circlearrowright 1308
\newsymbol\circlearrowleft 1309
\undefine\rightleftharpoons
\newsymbol\rightleftharpoons 130A
\newsymbol\leftrightharpoons 130B
\newsymbol\boxminus 120C
\newsymbol\Vdash 130D
\newsymbol\Vvdash 130E
\newsymbol\vDash 130F
\newsymbol\twoheadrightarrow 1310
\newsymbol\twoheadleftarrow 1311
\newsymbol\leftleftarrows 1312
\newsymbol\rightrightarrows 1313
\newsymbol\upuparrows 1314
\newsymbol\downdownarrows 1315
\newsymbol\upharpoonright 1316
 
\newsymbol\downharpoonright 1317
\newsymbol\upharpoonleft 1318
\newsymbol\downharpoonleft 1319
\newsymbol\rightarrowtail 131A
\newsymbol\leftarrowtail 131B
\newsymbol\leftrightarrows 131C
\newsymbol\rightleftarrows 131D
\newsymbol\Lsh 131E
\newsymbol\Rsh 131F
\newsymbol\rightsquigarrow 1320
\newsymbol\leftrightsquigarrow 1321
\newsymbol\looparrowleft 1322
\newsymbol\looparrowright 1323
\newsymbol\circeq 1324
\newsymbol\succsim 1325
\newsymbol\gtrsim 1326
\newsymbol\gtrapprox 1327
\newsymbol\multimap 1328
\newsymbol\therefore 1329
\newsymbol\because 132A
\newsymbol\doteqdot 132B
 
\newsymbol\triangleq 132C
\newsymbol\precsim 132D
\newsymbol\lesssim 132E
\newsymbol\lessapprox 132F
\newsymbol\eqslantless 1330
\newsymbol\eqslantgtr 1331
\newsymbol\curlyeqprec 1332
\newsymbol\curlyeqsucc 1333
\newsymbol\preccurlyeq 1334
\newsymbol\leqq 1335
\newsymbol\leqslant 1336
\newsymbol\lessgtr 1337
\newsymbol\backprime 1038
\newsymbol\risingdotseq 133A
\newsymbol\fallingdotseq 133B
\newsymbol\succcurlyeq 133C
\newsymbol\geqq 133D
\newsymbol\geqslant 133E
\newsymbol\gtrless 133F
\newsymbol\sqsubset 1340
\newsymbol\sqsupset 1341
\newsymbol\vartriangleright 1342
\newsymbol\vartriangleleft 1343
\newsymbol\trianglerighteq 1344
\newsymbol\trianglelefteq 1345
\newsymbol\bigstar 1046
\newsymbol\between 1347
\newsymbol\blacktriangledown 1048
\newsymbol\blacktriangleright 1349
\newsymbol\blacktriangleleft 134A
\newsymbol\vartriangle 134D
\newsymbol\blacktriangle 104E
\newsymbol\triangledown 104F
\newsymbol\eqcirc 1350
\newsymbol\lesseqgtr 1351
\newsymbol\gtreqless 1352
\newsymbol\lesseqqgtr 1353
\newsymbol\gtreqqless 1354
\newsymbol\Rrightarrow 1356
\newsymbol\Lleftarrow 1357
\newsymbol\veebar 1259
\newsymbol\barwedge 125A
\newsymbol\doublebarwedge 125B
\undefine\angle
\newsymbol\angle 105C
\newsymbol\measuredangle 105D
\newsymbol\sphericalangle 105E
\newsymbol\varpropto 135F
\newsymbol\smallsmile 1360
\newsymbol\smallfrown 1361
\newsymbol\Subset 1362
\newsymbol\Supset 1363
\newsymbol\Cup 1264
 
\newsymbol\Cap 1265
 
\newsymbol\curlywedge 1266
\newsymbol\curlyvee 1267
\newsymbol\leftthreetimes 1268
\newsymbol\rightthreetimes 1269
\newsymbol\subseteqq 136A
\newsymbol\supseteqq 136B
\newsymbol\bumpeq 136C
\newsymbol\Bumpeq 136D
\newsymbol\lll 136E
 
\newsymbol\ggg 136F
 
\newsymbol\circledS 1073
\newsymbol\pitchfork 1374
\newsymbol\dotplus 1275
\newsymbol\backsim 1376
\newsymbol\backsimeq 1377
\newsymbol\complement 107B
\newsymbol\intercal 127C
\newsymbol\circledcirc 127D
\newsymbol\circledast 127E
\newsymbol\circleddash 127F
\newsymbol\lvertneqq 2300
\newsymbol\gvertneqq 2301
\newsymbol\nleq 2302
\newsymbol\ngeq 2303
\newsymbol\nless 2304
\newsymbol\ngtr 2305
\newsymbol\nprec 2306
\newsymbol\nsucc 2307
\newsymbol\lneqq 2308
\newsymbol\gneqq 2309
\newsymbol\nleqslant 230A
\newsymbol\ngeqslant 230B
\newsymbol\lneq 230C
\newsymbol\gneq 230D
\newsymbol\npreceq 230E
\newsymbol\nsucceq 230F
\newsymbol\precnsim 2310
\newsymbol\succnsim 2311
\newsymbol\lnsim 2312
\newsymbol\gnsim 2313
\newsymbol\nleqq 2314
\newsymbol\ngeqq 2315
\newsymbol\precneqq 2316
\newsymbol\succneqq 2317
\newsymbol\precnapprox 2318
\newsymbol\succnapprox 2319
\newsymbol\lnapprox 231A
\newsymbol\gnapprox 231B
\newsymbol\nsim 231C
\newsymbol\ncong 231D
\newsymbol\diagup 201E
\newsymbol\diagdown 201F
\newsymbol\varsubsetneq 2320
\newsymbol\varsupsetneq 2321
\newsymbol\nsubseteqq 2322
\newsymbol\nsupseteqq 2323
\newsymbol\subsetneqq 2324
\newsymbol\supsetneqq 2325
\newsymbol\varsubsetneqq 2326
\newsymbol\varsupsetneqq 2327
\newsymbol\subsetneq 2328
\newsymbol\supsetneq 2329
\newsymbol\nsubseteq 232A
\newsymbol\nsupseteq 232B
\newsymbol\nparallel 232C
\newsymbol\nmid 232D
\newsymbol\nshortmid 232E
\newsymbol\nshortparallel 232F
\newsymbol\nvdash 2330
\newsymbol\nVdash 2331
\newsymbol\nvDash 2332
\newsymbol\nVDash 2333
\newsymbol\ntrianglerighteq 2334
\newsymbol\ntrianglelefteq 2335
\newsymbol\ntriangleleft 2336
\newsymbol\ntriangleright 2337
\newsymbol\nleftarrow 2338
\newsymbol\nrightarrow 2339
\newsymbol\nLeftarrow 233A
\newsymbol\nRightarrow 233B
\newsymbol\nLeftrightarrow 233C
\newsymbol\nleftrightarrow 233D
\newsymbol\divideontimes 223E
\newsymbol\varnothing 203F
\newsymbol\nexists 2040
\newsymbol\Finv 2060
\newsymbol\Game 2061
\newsymbol\mho 2066
\newsymbol\eth 2067
\newsymbol\eqsim 2368
\newsymbol\beth 2069
\newsymbol\gimel 206A
\newsymbol\daleth 206B
\newsymbol\lessdot 236C
\newsymbol\gtrdot 236D
\newsymbol\ltimes 226E
\newsymbol\rtimes 226F
\newsymbol\shortmid 2370
\newsymbol\shortparallel 2371
\newsymbol\smallsetminus 2272
\newsymbol\thicksim 2373
\newsymbol\thickapprox 2374
\newsymbol\approxeq 2375
\newsymbol\succapprox 2376
\newsymbol\precapprox 2377
\newsymbol\curvearrowleft 2378
\newsymbol\curvearrowright 2379
\newsymbol\digamma 207A
\newsymbol\varkappa 207B
\newsymbol\Bbbk 207C
\newsymbol\hslash 207D
\undefine\hbar
\newsymbol\hbar 207E
\newsymbol\backepsilon 237F


\count5=0
\count6=1
\count7=1
\count8=1
\count9=1

\def\references{\bigskip\noindent\hbox{\bf References}\medskip}
\def\remark{\medskip\noindent{\bf Remark.\ }}
\def\nextremark #1\par{\item{$\circ$} #1}
\def\firstremark #1\par{\bigskip\noindent{\bf Remarks.}
     \smallskip\nextremark #1\par}
\def\abstract#1\par{{\baselineskip=10pt
    \eightpoint\narrower\noindent{\eightbf Abstract.} #1\par}}
\def\equ(#1){\hskip-0.03em\csname e#1\endcsname}
\def\clm(#1){\csname c#1\endcsname}
\def\equation(#1){\eqno\tag(#1)}
\def\tag(#1){(\number\count5.
	      \number\count6)
    \expandafter\xdef\csname e#1\endcsname{
    (\number\count5.\number\count6)}
    \global\advance\count6 by 1}
\def\claim #1(#2) #3\par{
    \vskip.1in\medbreak\noindent
    {\bf #1\ \number\count5.\number\count7.\ }{\sl #3}\par
    \expandafter\xdef\csname c#2\endcsname{#1~\number\count5.\number\count7}
    \global\advance\count7 by 1
    \ifdim\lastskip<\medskipamount
    \removelastskip\penalty55\medskip\fi}
\def\section#1\par{\vskip0pt plus.1\vsize\penalty-40
    \vskip0pt plus -.1\vsize\bigskip\bigskip
    \global\advance\count5 by 1
    \message{#1}\leftline
     {\magtenbf \number\count5.\ #1}
    \count6=1
    \count7=1
    \count8=1
    \nobreak\smallskip\noindent}
\def\subsection#1\par{\vskip0pt plus.05\vsize\penalty-20
    \vskip0pt plus -.05\vsize\medskip\medskip
    \message{#1}\leftline{\tenbf
    \number\count5.\number\count8.\ #1}
    \global\advance\count8 by 1
    \nobreak\smallskip\noindent}
\def\addref#1{\expandafter\xdef\csname r#1\endcsname{\number\count9}
    \global\advance\count9 by 1}
\def\proofof(#1){\medskip\noindent{\bf Proof of \csname c#1\endcsname.\ }}

\def\rightheadline{\hfil}
\def\leftheadline{\sevenrm\hfil HANS KOCH\hfil}
\headline={\ifnum\pageno=\firstpage\hfil\else
\ifodd\pageno{{\fiverm\rightheadline}\number\pageno}
\else{\number\pageno\fiverm\leftheadline}\fi\fi}
\footline={\ifnum\pageno=\firstpage\hss\tenrm\folio\hss\else\hss\fi}

\let\cl=\centerline

\let\sss=\scriptscriptstyle

\def\integer{{\Bbb Z}}

\def\real{{\Bbb R}}

\def\torus{{\Bbb T}}

\def\defeq{\mathrel{\mathop=^{\rm def}}}
\def\half{{1\over 2}}

\def\FF{{\cal F}}
\def\GG{{\cal G}}

\def\OO{{\cal O}}

\def\SS{{\cal S}}

%
\input miniltx

\ifx\pdfoutput\undefined
  \def\Gin@driver{dvips.def}  
\else
  \def\Gin@driver{pdftex.def} 
\fi
 
\input graphicx.sty
\resetatcatcode

\input texdraw
\font\foo=rphvr at 8 pt
%
\def\rename#1{color-#1}       
%
\def\ssLambda{{\sss\Lambda}}

\def\ELL{{\rm L}}
\def\defeq{\mathrel{\mathop=^{\rm def}}}
\def\placetext (#1 #2 #3){
\rmove (#1 #2)
\htext {#3}
\rmove (-#1 -#2)
}
\def\tile(#1 #2 #3 #4){
\drawdim {cm}
\placetext (0.42 0.10 #4)
\placetext (0.73 0.40 #2)
\placetext (0.42 0.68 #3)
\placetext (0.08 0.40 #1)
\rlvec( 1  0)
\rlvec( 0  1)
\rlvec(-1  0)
\rlvec( 0 -1)
\lfill f:0.95
\rmove (1 0)
}
\addref{Janot}
\addref{Senechal}
\addref{RadinA}
\addref{Mackay}
\addref{LevineSteinhardt}

\addref{RuelleA}
\addref{RadinB}
\addref{LeuzziParisi}
\addref{Miekisz}
\addref{GrunbaumShephard}
\addref{RuelleB}
\addref{Schrader}
\addref{RuelleC}
\addref{Simon}
\addref{Nevo}
\addref{CornfeldFominSinai}
\addref{Data}
\def\leftheadline{\fiverm\hfil HANS KOCH and CHARLES RADIN\hfil}
\def\rightheadline{\sevenrm\hfil Quasicrystals at
positive temperature\hfil}
\cl{{\magtenbf Modelling quasicrystals at}}
\cl{{\magtenbf positive temperature}}
\bigskip

\cl{Hans Koch
\footnote{$^1$}
{{\sevenrm Department of Mathematics,
The University of Texas at Austin, Austin, TX 78712}}
and Charles Radin~$^{1,}\!\!$
\footnote{$^2$}
{{\sevenrm Research supported in part by NSF Grant DMS-0700120}}}
\vskip.5truein

\abstract
We consider a two-dimensional lattice model of equilibrium 
statistical mechanics, using nearest neighbor interactions based on
the matching conditions for an aperiodic set of $16$ Wang tiles.
This model has uncountably many ground state
configurations, all of which are nonperiodic.
The question addressed in this paper is
whether nonperiodicity persists at low but positive temperature.
We present arguments, mostly numerical, that this is indeed the case.
In particular, we define an appropriate order parameter,
prove that it is identically zero at high temperatures,
and show by Monte Carlo simulation that it is
nonzero at low temperatures.

\vskip.5truein
\section Introduction

Certain alloys are believed to exhibit, at low temperature, a state of
thermal equilibrium which is solid but not crystalline (as determined
for instance by X-ray diffraction), a state called quasicrystalline [\rJanot].

In part to explain the observed
diffraction patterns, it is common to model the energy ground state of such a 
material by
using (aperiodic) tilings [\rJanot, \rSenechal, \rRadinA]. But ever since such models 
have been proposed
[\rMackay, \rLevineSteinhardt] there has been the need to determine whether 
they actually are
useful to explain the behavior of materials at positive temperature,
that is, it is unclear whether such models exhibit a phase transition
from the usual disorder at high temperature to an aperiodically ordered state at low
but positive temperature.

We will analyze a two-dimensional lattice ``tiling'' model with
appropriate energy ground states in an attempt to make some progress
in this problem. The lattice model is of a standard form
[\rRuelleA, \rRadinB].
In fact the specific model has been discussed already,
by Leuzzi and Parisi [\rLeuzziParisi],
though they concentrated on the degeneracy of the energy ground state
of the model, and a possible connection with the nonequilibrium
behavior of glasses and similar materials. See also [\rMiekisz].

We will not be giving a proof of a phase transition in this
model. Indeed, there are very few models for which one can prove phase
transitions. A proof of a quasicrystalline phase may well require a
new basic technique. But at least it would be useful to have a good
order parameter, and to develop some intuition about the nature of the
order, on which to base a future argument for a transition. 
This is our goal. We will introduce such an order
parameter for our model, prove that it vanishes identically at high
temperature, and present numerical evidence that it is
nonzero below some critical value of the temperature.

\vfil\eject

\section The model and main results

\subsection An overview

We consider a model
of interacting particles on the lattice $\integer^2$,
where each site can be in one of $16$ different states.
The possible one-particle states are identified
with Ammann's $16$ ``prototiles'' depicted in Fig.~1 below.
Each of these prototiles is a unit square in $\real^2$,
centered at the origin, whose edges each carry a label
from one of the sets $S=\{1,2\}$ or $L=\{3,4,5,6\}$.
(Rotations are not allowed.)
Notice that the horizontal edges of a prototile
are either both of type $S$, or both of type $L$.
Similarly for the vertical edges.

\bigskip
\centertexdraw{
\tile(2 1 1 2)
\rmove(0.3 0)
\tile(1 1 4 6)
\rmove(0.3 0)
\tile(1 2 4 5)
\rmove(0.3 0)
\tile(1 2 6 3)
\rmove(0.3 0)
\tile(2 2 5 3)
\rmove(0.3 0)
\tile(6 4 1 1)
\rmove(0.3 0)
\tile(5 4 2 1)
\rmove(0.3 0)
\tile(3 6 2 1) \move (0 -1.3)
\tile(3 5 2 2)
\rmove(0.3 0)
\tile(5 3 3 5)
\rmove(0.3 0)
\tile(5 5 3 6)
\rmove(0.3 0)
\tile(6 3 5 5)
\rmove(0.3 0)
\tile(6 5 5 6)
\rmove(0.3 0)
\tile(6 6 5 4)
\rmove(0.3 0)
\tile(4 5 6 6)
\rmove(0.3 0)
\tile(4 6 6 4)
\rmove(0.3 0)
}
\smallskip
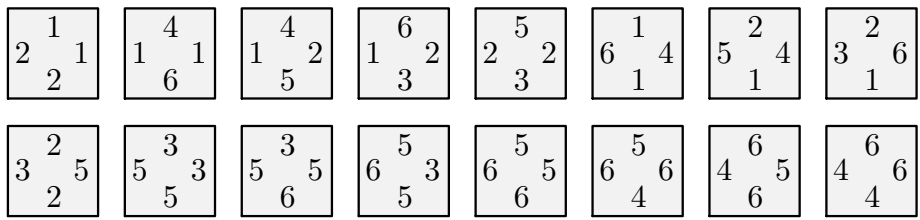
\centerline{\eightpoint {\bf Fig.~1.} Ammann's $16$ prototiles.}

\bigskip\noindent
Denoting Ammann's set of prototiles by $A$,
a configuration of our particle system
is given by a map $\sigma:j\in \integer^2\mapsto\sigma_j\in A$.
Such a configuration $\sigma$ can also be represented by
a collection of tiles $\{j+\sigma_j: j\in\integer^2\}$,
where $j+\sigma_j$ is a translated copy of the prototile $\sigma_j\,$.
This collection
defines a covering of $\real^2$ by
squares (with labeled edges) whose centers lie on $\integer^2$,
and whose interiors are mutually disjoint.
In such a covering,
any pair of tiles $\{i+\sigma_i,j+\sigma_j\}$
that share a common edge assigns two labels to this edge.
If the two labels disagree, then the set $\{i,j\}$
is referred to as a {\it defect} of $\sigma$.
A configuration or covering without defects will be called an $A$-tiling,
or tiling for short.

It is shown in [\rGrunbaumShephard] that such tilings exist,
and that they are all nonperiodic.
The nonperiodicity follows from the existence of an inflation rule,
which is related to the substitution rule
$S\rightarrow L$ and $L\rightarrow SL$ for Fibonacci sequences.
(Fibonacci sequences are $2$-sided-infinite words, obtained from the sequence
$$
L\rightarrow SL\rightarrow LSL\rightarrow SLLSL
\rightarrow LSLSLLSL
\rightarrow
SLLSLLSLSLLSL\rightarrow\ldots
$$
via translations and limits.)
More precisely, all tiles in a given row of an $A$-tiling
have the same vertical type ($S$ or $L$),
and their horizontal type defines a Fibonacci sequence.
Similarly for the columns.
Furthermore, the inflation map is invertible
on the set $\GG$ of all $A$-tilings.
Besides nonperiodicity,
this also implies that $\GG$ carries a
unique translation-invariant probability measure $\lambda$ [\rRadinA].

Since all $A$-tilings are nonperiodic,
no finite ``patch'' determines a tiling uniquely.
However, large patches of a tiling are strongly correlated,
even if they are arbitrarily far apart.
The question that motivated our analysis
is whether this ``long range order'' is still present
in a thermodynamic ensemble, where typical tile configurations
have a small but positive density of defects.
A thermodynamic ensemble is a probability measure
on the space $\SS$ of tile configurations
$\sigma:\integer^2\to A$, indexed by an inverse temperature $\beta>0$.
A more precise definition will be given below.
Roughly speaking, our measure $\nu_{\beta}$
assigns a relative weight $\exp[-\beta H(\sigma)]$
to a tile configuration $\sigma$, where $H(\sigma)$
is the number of defects of $\sigma$.
The minimizers of $H$, also referred to as ground state configurations,
are precisely the $A$-tilings.
(It is not difficult to prove that any
state obtained as a limit $\beta\to\infty$ gives full measure to
these configurations [\rRuelleB, \rSchrader].)

To test for long range order,
we only consider the type ($S$ or $L$) of the four tile edges.
This identification by type
defines an equivalence relation ``$\sim$'' on $A$, with four
equivalence classes.
Given a configuration $\sigma$, an $A$-tiling $\gamma$,
and a finite region $\Lambda\subset\integer^2$,
denote by $\phi_\ssLambda(\sigma,\gamma)$
the fraction of sites $j\in\Lambda$ where $\sigma_j\sim\gamma_j$.
The limit as $\Lambda\uparrow\integer^2$, if it exists,
will be denoted by $\bar\phi(\sigma,\gamma)$, and we define
$\bar\psi(\sigma)=\int_\GG\bar\phi(\sigma,\gamma) d\lambda(\gamma)$.
One of our goals is to study the quantity
$$
Q_\beta(\gamma)=\int_\SS
{\bar\phi(\sigma,\gamma)\over\bar\psi(\sigma)}\,d\nu_{\beta}(\sigma)\,,
\qquad \gamma\in\GG\,,
\equation(OP)
$$
which measures how much
a typical tile configuration at inverse temperature $\beta$
aligns with the ground state configuration $\gamma$.
In the absence of any preference,
$Q_\beta$ is the constant function $1$,
or equivalently, the ``order parameter''
$$
q(\beta)=\int_\GG\!Q_\beta(\gamma)
\ln[Q_\beta(\gamma)] d\lambda(\gamma)
\equation(op)
$$
is identically zero.
We will prove that this is the case for sufficiently small $\beta>0$.
This result is independent of the choice of boundary condition
in defining $\nu_\beta$ through the energy function $H$.

\vskip 0.2in
\hbox{
\hskip0.04in
\includegraphics[height=2.5in, width=2.5in, angle=-90]{\rename{order_parameter}}
\includegraphics[height=2.5in, width=2.5in, angle=-90]{\rename{boundary_overlap}}}
\noindent
\hbox{\eightpoint
\phantom{xxxxxx}
Fig.~2 \ The order parameter $q_\beta\,$.
\phantom{xxxxxxxxxxxxxx}
Fig.~3 \ The overlap $Q_\beta(\tau)$.
\hfil}

\vskip 0.2in
Our remaining results are purely numerical.
In what follows, we choose as boundary conditions (at infinity)
a fixed $A$-tiling $\tau$.
It is important to keep this in mind.
Fig.~2 shows the values of $q(\beta)$
obtained via Monte Carlo simulations,
for tile configurations of size $N\times N$,
with $N$, a power of $2$, ranging from $32$ to $256$.
These results suggest that $q(\beta)$
becomes negative as $\beta$ is increased
past a critical value $\beta_c\approx 2.4$.
If correct, this would imply the existence of a phase transition,
from a disordered state for $\beta<\beta_c$
to an ordered state for $\beta>\beta_c\,$,
where translation invariance is broken.
As one would expect,
$Q_\beta(\gamma)$ takes its maximum at $\gamma=\tau$.
Fig.~3 shows the simulated values of $Q_\beta(\tau)$.

Similar signs of an order-disorder transition were found
in [\rLeuzziParisi],
using Monte Carlo simulations for $8\le N\le 32$,
with free (but eventually frozen) boundary conditions.
At these values of $N$,
there is evidence that the phase transition is of second order,
with a power-law or logarithmic divergence
of the specific heat (as $\beta\to\beta_c$),
depending on the model used to fit the data.
Our numerical results for $32\le N\le 256$
clearly favor the second alternative, if either.
In fact, we find a slowdown in the increase
of the specific heat, suggesting that
the phase transition is of third or higher order.
Our simulated values for the energy per tile,
and for the specific heat, are shown in Figs. 4 and 5.
The curve labeled {\foo dE/dT} is
the (discrete) derivative of the energy, for $N=256$,
and the other three curves in Fig.~5
were obtained from the energy fluctuations.

\vskip 0.1in
\hbox{
\hskip0.04in
\includegraphics[height=2.5in, width=2.5in, angle=-90]{\rename{energy_density}}
\includegraphics[height=2.5in, width=2.5in, angle=-90]{\rename{specific_heat}}}
\noindent
\hbox{\eightpoint
\phantom{xxxx}
Fig.~4 \ The average energy per tile.
\phantom{xxxxxxxxx}
Fig.~5 \ The specific heat.
\hfil}
\vskip 0.2in

Fig.~6 shows a two-point correlation $C_\beta(d)$
as a function of the (vertical) separation $d$.
For $\beta>\beta_c\,$, the correlation
approach a nonzero constant that depends on $\beta$,
indicating again the existence of long range order at low temperature.
The correlation length $C_\beta^{-1}(1/4)$ is shown in Fig.~7.
It diverges roughly like $(\beta_c-\beta)^{-7}$
as $\beta\uparrow\beta_c\,$.
Such power law behavior is again similar to what is observed
in models with ordered low temperature phases,
except that the exponent $7$ is unusually large.
By contrast, the correlations for $\beta\approx\beta_c$
decay roughly like $\exp(-cd^{1/2})$,
at least in the observed range.
This suggests that there is no
renormalization fixed point (nontrivial scaling limit)
associated with the critical value $\beta_c\,$.
Instead, there seems to be a ``line of fixed points''
for $\beta>\beta_c\,$.
This is a feature that is better known in models
with a continuous (internal) symmetry, such as the $XY$-model.

\vskip 0.2in
\hbox{
\hskip0.04in
\includegraphics[height=2.5in, width=2.5in, angle=-90]{\rename{correlations}}
\includegraphics[height=2.5in, width=2.5in, angle=-90]{\rename{correlation_length}}}
\noindent
\hbox{\eightpoint
\phantom{xx}
Fig.~6 \ Correlation: $C_\beta(d)$ versus $d$.
\phantom{xxxxxxxxxxx}
Fig.~7 \ Correlation length $C_\beta^{-1}(1/4)$.
\hfil}
\vskip 0.2in

\subsection Further observations

We now give a description of the ground states,
that can be used to compute and visualize the function $Q_\beta\,$.
It should also provide some insight into the
behavior of the model at low temperature.

Without the normalizing factor $\bar\psi^{-1}$
in the integral \equ(OP), our function $Q_\beta$ is analogous
to the order parameters used in models that have
periodic ground states and/or a compact internal symmetry group.
By being a function on $\GG$,
it is implicitly covariant under any symmetry of the model,
including translations.
If none of the symmetries are broken,
then the corresponding ``entropy'' $q(\beta)$ vanishes.
Thus, $Q_\beta$ seems to be a natural order parameter
in models with a large number of ground states.
For the tiling model considered here, this number is uncountable.
Nevertheless, $Q_\beta$ is easy to compute.
The reason is that, as we will see below,
$\GG/\!\!\sim$ can be identified with the torus $\torus^2$,
where translations $(T^j\sigma)_i=\sigma_{i+j}$
act by irrational rotations.
Each equivalence class $[a]=\{x\in A: x\sim a\}$
of $A$ corresponds to one of four
disjoint rectangles in a covering of $\torus^2$,
and $\psi_\ssLambda^{-1}(\sigma)\phi_\ssLambda(\sigma,\gamma)$
is the probability that
the rectangle $[\sigma_j]$ contains the point $T^j\gamma$,
for a randomly chosen site $j\in\Lambda$.

To be more precise,
given any $X,Y\in\{S,L\}$,
denote by $X\times Y$ the set (equivalence class)
of all prototiles in $A$ whose horizontal and vertical
edges are of type $X$ and $Y$, respectively.
Let $R=\{S\!\times\!S, S\!\times\!L, L\!\times\!S, L\!\times\!L\}$.
Then to every tile configuration $\sigma:\integer^2\to A$
we can associate a function $[\sigma]:\integer^2\to R$,
by setting $[\sigma]_j=[\sigma_j]$.
If $\gamma$ is an $A$-tiling, then $[\gamma]$
is a product of two Fibonacci sequences
$k\mapsto x_k$ and $k\mapsto y_k\,$,
in the sense that $[\gamma]_j=x_{j_1}\times y_{j_2}\,$.
Conversely, any product of two Fibonacci sequences
can be obtained in this way.
Thus, the set of all such $R$-tilings $[\gamma]$
is the product $\FF\times\FF$,
where $\FF$ denotes the set of all Fibonacci sequences.
Since $Q_\beta(\gamma)$ only depends
on the equivalence class $[\gamma]$,
it suffices to find a convenient description
for the Fibonacci sequences.
(In fact, $[\gamma]=\{\gamma\}$ for almost every tiling $\gamma$,
but we will not use this here.)

One such description is the following [\rSenechal, p.~128].
Given a real number $\vartheta$,
and a partition $\{J(S),J(L)\}$ of the circle $\torus=\real/\integer$,
we can associate with any angle $\alpha\in\torus$
a sequence $x\in\{S,L\}^{\integer}$ by setting
$$
x_k=\cases{
S\,, &if $\alpha+k\vartheta\in J(S)$;\cr
L\,, &if $\alpha+k\vartheta\in J(L)$.\cr}
\equation(AngleToFibo)
$$
Let now $\vartheta$ be the inverse golden mean, $\vartheta=\half(\sqrt{5}-1)$.
The Fibonacci sequences are obtained by choosing
either $J(L)=[0,\vartheta)$ and $J(S)=[\vartheta,1)$,
or else $J(L)=(0,\vartheta]$ and $J(S)=(\vartheta,1]$.
The two sets of sequences differ only by a countable set,
corresponding to angles $\alpha+k\vartheta$ that are zero (modulo $1$).
This set has measure zero, so we can ignore it.
Thus, for our purposes, $\FF$ can be identified with
the circle $\torus$.

In this representation, translations $(T^k x)_m=x_{m+k}$
on $\FF$ become irrational rotations $R^k\alpha=\alpha+k\vartheta$
on the circle.
Similarly, $\integer^2$-translations on $\FF\times\FF$
are represented by irrational rotations
$R^j\alpha=(\alpha_1+j_1\vartheta,\alpha_2+j_2\vartheta)$
on the torus $\torus^2$.
In both cases, the unique invariant measure is Lebesgue measure.
We will describe later how these properties can be
used for numerical computations.

While a ground state configuration $\gamma$
determines a point $\alpha$ on the torus,
an $n\times n$ ``patch'' of $\gamma$
determines a rectangular neighborhood of $\alpha$
of linear size $\OO(n^{-1})$.
Thus, a typical low temperature configuration
determines locally a pair of ``fuzzy'' angles.
Unlike in the $XY$-model, the energy associated
with a gradual change of the column (row) angle
over a horizontal (vertical) distance $d$ does not decrease with $d$.
This follows from the fact that the density of
letter-mismatches between two Fibonacci sequences
is asymptotically proportional to the angle difference.
A similar argument may apply in the other directions,
based on the ``slanted'' Fibonacci sequences described in the remark below.
Thus, it seems plausible that the model can maintain
long range order at low temperatures,
despite the fact that there are uncountably many ground states.

By analogy with the $XY$-model,
one might ask about the existence of vortices and/or dipoles
in our tiling model.
The question is meaningful only at reasonably low temperature,
since the angles are ill defined at high temperature.
So isolated vortices are unlikely to play a major role.
But in a model with slowly varying angles,
a dipole can be associated with two successive crossings
through a fixed value.
Such dipoles (horizontal and vertical) are a prominent feature
at temperatures near $\beta_c\,$.
But they are rarely isolated and thus hard to analyze systematically.
Besides these horizontal/vertical dipoles,
one can also observe ``slanted'' dipoles
whose ends are single defects (not type mismatches),
and whose connecting line has a slope near $\vartheta^{\pm 1}$.
These slanted dipoles seem to be the main source
of entropy at very low temperatures.
Their density has no visible singularity over the range of
temperatures considered,
so they do not appear to play a major role
in the observed phase transition.

\remark
In a different representation of the Ammann tilings [\rGrunbaumShephard],
the $16$ prototiles are rectangles,
whose $L$-edges and $S$-edges
have lengths $\vartheta$ and $1-\vartheta$, respectively.
In this representation,
it is possible to replace the numeric edge-labels
by two types of line segments, say blue and green, transverse to the edges, 
such that a perfect tiling is characterized
by the blue (and similarly the green) segments
combining into a parallel sequence of straight lines,
known as Ammann bars.
The bars are slanted, with slopes $\vartheta^{\pm 1}$,
and it should not be too surprising
that the spacings between the blue
(as well as the green) bars define a Fibonacci sequence.
The same can be done with square tiles,
except that the bars are only piecewise linear.
This shows that the matching rules for the Ammann tiles
enforce, primarily and in a direct way,
products of Fibonacci sequences.

\subsection The model

In this section, we give a more detailed description of the model
and show that the order parameter \equ(op) vanishes
for small $\beta>0$.
We recall that our simulations were carried out
with boundary conditions given by a tiling $\tau$.
For the purpose of this section,
$\tau$ could be any configuration in $\SS$.
Thus, we shall suppress the dependence on $\tau$ in our notation.

We start by considering finite regions $\Lambda\subset\integer^2$.
The energy $H_\ssLambda(\sigma)$ of a configuration
$\sigma\in\SS$ is defined
as the number of defects of $\sigma$ that intersect $\Lambda$.
Given a real number $\beta>0$,
and a finite subset $\Lambda$ of $\integer^2$,
the Gibbs state (for $\Lambda$) at temperature $1/\beta$,
with boundary condition $\tau$,
is the functional that assigns to a continuous function $f$
(for the product topology) on $\SS$
the value
$$
\langle f\rangle_{\beta,\Lambda}
=Z_{\beta,\Lambda}^{-1}\sum_{\sigma\in\SS_{\Lambda}}
f(\sigma)e^{-\beta H_\Lambda(\sigma)}\,.
\equation(lfl)
$$
Here, $\SS_{\ssLambda}$ is the set of configurations
$\sigma\in\SS$ that agree with $\tau$ outside $\Lambda$,
and $Z_{\beta,\Lambda}$ is a normalization constant,
determined by the condition $\langle 1\rangle_{\beta,\Lambda}=1$.
Taking a limit $\Lambda\uparrow\integer^2$
along squares
defines a Gibbs measure $\nu_{\beta}$ on $\SS$,
$$
\int_\SS f\,d\nu_{\beta}=
\langle f\rangle_{\beta}\defeq
\lim_{\Lambda\uparrow\integer^2}\langle f\rangle_{\beta,\Lambda}\,.
\equation(lflimit)
$$
By well known results in the theory of lattice models [\rRuelleC, \rSimon],
the measure $\nu_{\beta}$ for small positive $\beta$
is translation invariant, ergodic, and mixing.
In particular, $\nu_{\beta}$ does not depend on the choice of
boundary condition $\tau$.
(For large $\beta$ the limit may have to be taken
along subsequences, and it can depend on $\tau$.)

Let $\nu$ be any translation-invariant probability measure on $\SS$,
and consider the space $\Omega=\SS\times\GG$,
equipped with the product measure $\mu=\nu\times\lambda$.
This measure is invariant under translations
$T^j(\sigma,\gamma)=(T^j\sigma,T^j\gamma)$.
Thus by the ergodic theorem [\rNevo], if $\phi$ is any function
in $\ELL^1(\Omega)$, then the orbit averages
$$
\phi_n(\sigma,\gamma)=
{1\over 4n^2}\sum_{j_1=-n}^{n-1}\sum_{j_2=-n}^{n-1}
\phi\bigl(T^j(\sigma,\gamma)\bigr)
\equation(orbav)
$$
converge $\mu$-almost everywhere to
a function $\bar\phi$ in $\ELL^1(\Omega)$, as $n\to\infty$.
In what follows, let
$\phi(\sigma,\gamma)=\theta(\sigma_0\sim\gamma_0)$,
where $\theta({\rm true})=1$ and $\theta({\rm false})=0$.
Now assume that $\nu$ is mixing.
Then, by a standard result in ergodic theory [\rCornfeldFominSinai,
p.~228],
the measure $\mu$ is ergodic.
As a result, $\bar\phi$ is the constant function
with value $\int_\Omega\phi\,d\mu$.

This shows that the function $Q_\beta$
defined in \equ(OP) is well defined,
as long as the Gibbs measure $\nu_\beta$ is translation invariant.
Furthermore, if $\nu_\beta$ is mixing, then this function
is identically $1$, and $q(\beta)=0$.
As mentioned above, this holds for sufficiently small $\beta>0$.

We note that these arguments do not show that $Q_\beta$
is well defined for all values of $\beta$,
although this seems likely to be true.
(It is true for instance at zero temperature, as we will see later.)
In any case, if translation invariance is broken at some $\beta>0$,
then a phase transition has to occur.

The correlations described earlier are given by
$$
\Gamma_\beta(i)=\langle V_0V_i\rangle_\beta
-\langle V_0\rangle_\beta\langle V_i\rangle_\beta\,,
\qquad i\in\integer^2\,,
\equation(Cbetai)
$$
where $V_i=V\circ T^i$,
and where $V:\SS\to\real$ is defined as follows.
Denote by $x_j(\sigma)$ the horizontal type ($L$ or $S$)
of the tile $\sigma_j$ in a configuration $\sigma$.
Define $f(S)=-1$ and $f(L)=1$.
Then $V(\sigma)$ is the value of $f(x_j(\sigma))$,
averaged over all sites $j$
in some fixed finite region containing the origin.
The data in Fig.~6 are only for the correlation
$C_\beta(d)=\Gamma_\beta(i)$ in the vertical direction $i=(0,d)$.
The correlations for $V$ in other directions
have been computed as well, but they are not shown here.
They have large oscillations (related to the Fibonacci sequence)
and are less convenient for estimating a correlation length.
We also considered a purely probabilistic measure of correlations,
namely the relative entropy of the random variable
$(j,\sigma)\mapsto x_j(\sigma)$ and its translates.
The results are not qualitatively different from those
found via \equ(Cbetai).

\section Computations

Using the correspondence between ground states
(modulo equivalence) and points on the torus,
the overlap \equ(orbav) of a configuration $\sigma$
with a tiling $\gamma$ can now be written as follows.
Let $\alpha$ be the point on $\torus^2$ defined by $\gamma$.
Writing $[\sigma_j]=x_j\times y_j\,$, we have
$$
\phi_n(\sigma,\gamma)=
{1\over 4n^2}\sum_{j_1=-n}^{n-1}\sum_{j_2=-n}^{n-1}
\chi(J(x_j),\alpha_1+j_1\vartheta)
\chi(J(y_j),\alpha_2+j_2\vartheta)\,,
\equation(phin)
$$
where $b\mapsto\chi(B,b)$ denotes the indicator function
of a set $B\subset\torus$.
The integral of $\phi_n$ over $\alpha\in\torus^2$ is given by
$$
\psi_n(\sigma)=
{1\over 4n^2}\sum_{j_1=-n}^{n-1}\sum_{j_2=-n}^{n-1}
|J(x_j)| |J(y_j)|\,.
\equation(psin)
$$

As was shown earlier, $\psi_n^{-1}\phi_n\to 1$ as $n\to\infty$,
for small $\beta$.
The limit can also be computed
at zero temperature.
In this case, the double sum in \equ(phin) factorizes
into a product of simple sums.
Without loss of generality (due to translation invariance),
we can assume that $\sigma$ corresponds to the torus point $0$.
By using the ergodicity of irrational rotations, one finds that
$\phi_n\to\bar\phi$ a.e.~on $\GG\times\GG$, with
$\bar\phi(\sigma,\gamma)=\varphi(\alpha_1)\varphi(\alpha_2)$,
where $\varphi$ is the piecewise linear function
$$
\varphi(t)=\cases{
1-2|t|\,,       &if $|t|<1-\vartheta$;\cr
2\vartheta-1\,, &otherwise.\cr}
$$
The functions $\psi_n$ converge a.e.~to the constant $\kappa^2$,
where $\kappa=\vartheta^2+(1-\vartheta)^2$.

We expect that
$Q_\beta(\gamma)\approx\kappa^{-2}\varphi(\alpha_1)\varphi(\alpha_2)$
for large values of $\beta$.
This is indeed observed numerically,
but this is to be expected in a finite system.
For comparison, we show in Fig.~8 the computed values
of $Q_\beta(\gamma)$, as a function of the point $\alpha\in\torus^2$,
for the inverse temperatures $2.3$ and $2.4$.
(The observed transition for $N=256$ is between these values.)

\vskip 0.2in
\hbox{
\hskip 0.0in
\includegraphics[height=5.0in, width=3.0in, angle=-90]{\rename{distribution}}}
\noindent
\hbox{\eightpoint
\phantom{xxxxxxxxxxxxxx}
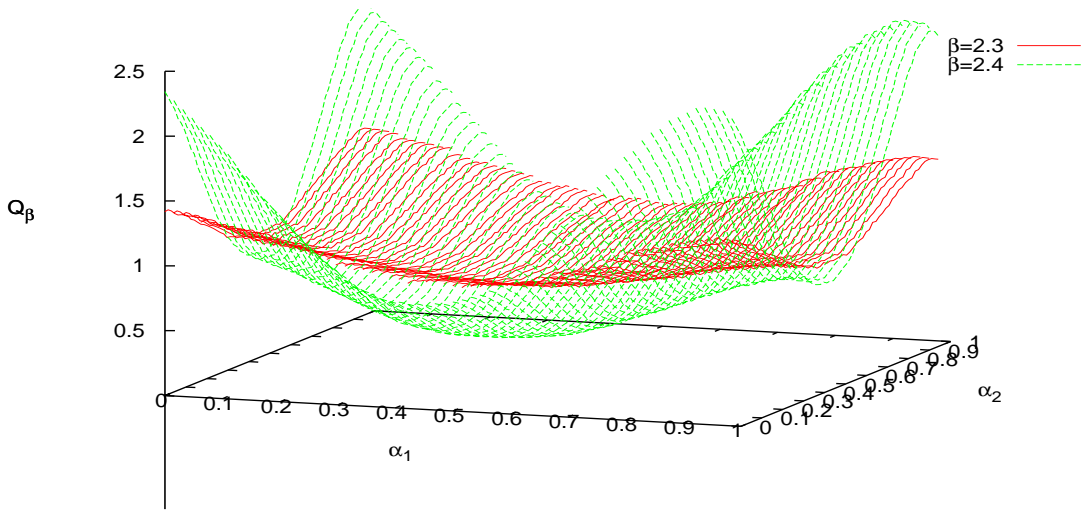
Fig.~8 \ Overlap $Q_\beta$ with the different ground states.
\hfil}

\bigskip\noindent
We recall that our simulations were carried out
with boundary conditions given by a tiling $\tau$.
This is why (and where) the graphs in Fig.~8 have a single maximum.
If we had used other boundary conditions,
then the limit of $\sigma_\beta$ as $\beta\to\infty$
would be a mixture of pure tiling states, possibly very complicated,
and the system might resemble a spin glass,
as was observed for instance in [\rLeuzziParisi].
In this context, we should mention that
the number of defect-free configurations
on an $N\times N$ lattice square
is bounded by $e^{cN}$, as was already described in [\rLeuzziParisi].
The number of such configurations that can be extended to a full
$A$-tiling is only $\OO(N^2)$.

In our numerical computations, we evaluate the sum in \equ(phin)
on a finite $256\times 256$ grid of points $\alpha\in\torus^2$.
(Choosing a finer grid gives no significant improvement.)
Averaging $\phi_n$ over $\alpha$ then yields $\psi_n\,$.
This is done for each individual tile configuration $\sigma$
in a collection $\Sigma_{2n,\beta}$, obtained via Monte Carlo simulation.
Then we perform the integral \equ(OP)
by averaging $\psi_n^{-1}\phi_n$ over the configurations
in $\Sigma_{2n,\beta}\,$.

Each of our ensembles $\Sigma_{N,\beta}$ contains $10^3$ configurations.
For increased flexibility, they were computed beforehand
and stored for analysis later [\rData].
The configurations in $\Sigma_{N,\beta}$ are separated
from each other by at least $M_{N,\beta}$ Monte-Carlo steps
(updates of individual tiles),
where $M_{N,\beta}$ was determined by monitoring overlaps
with the appropriate starting configuration,
and settling times for various observables,
to eliminate any visible dependence or bias.
To give an example, $M_{256,2.45}\approx 2.3*10^{12}$.
The initial configurations for $N=256$ were obtained
by slowly cooling a random configuration.
Patches of the resulting configurations were also used
to generate the starting points for $N<256$.

\references

{\ninepoint

\item{[\rJanot]} 
C.~Janot,
{\it Quasicrystals: A Primer},
Oxford University Press (1997).

\item{[\rSenechal]}
{M.~Senechal},
{\it Quasicrystals and geometry},
Cambridge University Press (1995).

\item{[\rRadinA]}
C.~Radin,
{\it Miles of Tiles},
American Mathematical Society (1999).

\item{[\rMackay]}
A.~Mackay, 
{\it Crystallography and the Penrose pattern},
Physica {\bf 114A}, 609--613 (1982).

\item{[\rLevineSteinhardt]}
D.~Levine, P.J.~Steinhardt, 
{\it Quasicrystals: a new class of ordered structures},
Phys.~Rev.~Lett. {\bf 53}, 2477--2480 (1984).

\item{[\rRuelleA]}
D.~Ruelle,
{\it Thermodynamic Formalism},
Addison-Wesley (1978).

\item{[\rRadinB]}
C. Radin, 
{\it Tiling, periodicity, and crystals}, 
J.~Math.~Phys. {\bf 26}, 1342--1344 (1985).

\item{[\rLeuzziParisi]} 
L.~Leuzzi, G.~Parisi,
{\it Thermodynamics of a tiling model},
J.~Phys.~A {\bf 33}, 4215--4225 (2000).

\item{[\rMiekisz]}
J.~Mi\c ekisz,
{\it Many phases in systems without periodic ground states},
Commun.~Math.~Phys. {\bf 107}, 577--586 (1986).

\item{[\rGrunbaumShephard]}
B.~Gr\"unbaum, G.C.~Shephard,
{\it Tilings and Patterns},
Freeman (1986).

\item{[\rRuelleB]} 
D.~Ruelle,
{\it Some remarks on the ground state of infinite
systems in statistical mechanics}, 
Commun.~Math.~Phys. {\bf 11}, 339--345 (1969).

\item{[\rSchrader]}
R.~Schrader, 
{\it Ground states in classical lattice systems with
hard core}, 
Commun.~Math.~Phys. {\bf 16}, 247--264 (1970).

\item{[\rRuelleC]} D.~Ruelle,
{\it Statistical mechanics, rigorous results},
W.A.~Benjamin Inc. (1969).

\item{[\rSimon]} B.~Simon,
{\it The statistical mechanics of lattice gases, Volume 1},
Princeton University Press (1993).

\item{[\rNevo]}
A.~Nevo,
{\it Pointwise ergodic theorems for actions of groups},
in Handbook of Dynamical Systems, IB, 
ed. A. Katok, B. Hasselblatt,
Elsevier (2005).

\item{[\rCornfeldFominSinai]}
I.~Cornfeld, S.~Fomin, Ya.~Sinai, 
{\it Ergodic Theory},
Springer-Verlag (1982).

\item{[\rData]}
The numerical ensembles $\Sigma_{N,\beta}$
are available from the authors upon request.

}
\bye